\documentclass[fleqn,10pt]{wlscirep}
\usepackage[utf8]{inputenc}
\usepackage[T1]{fontenc}
\usepackage{amsmath,amssymb}
\usepackage{subcaption}
\usepackage{graphicx}

\title{Reentrant Localization Length in a Quasiperiodic Thue-Morse Chain}

\author[1]{Taylan Y\i ld\i z}
\author[1,*]{B. Tanatar}
\affil[1]{Department of Physics, Bilkent University, Ankara, 06800, T\"urkiye}
\affil[*]{email:tanatar@fen.bilkent.edu.tr}

\begin{abstract}
We investigate single-particle localization in a dimerized
Su--Schrieffer--Heeger chain whose quasiperiodic onsite potential is masked by
the deterministic Thue--Morse sequence.  The behavior is characterized by the transfer-matrix Lyapunov exponent, and by the participation moments of exactly diagonalized chains.  For intracell-dominated
hopping the mid-spectrum Lyapunov exponent depends nonmonotonically on the
potential strength, developing a pronounced minimum that enhances the
localization length by an order of magnitude over an intermediate window,
whereas an intercell-dominated chain responds monotonically.  The exponent
remains positive at every sampled nonzero potential strength and at every state
of the spectral sector, and the second participation moment saturates over the
full range of system sizes, so the intermediate window is an enhancement of a
finite localization length and not a change of localization class.  Finite-size generalized dimensions follow the same response in real space and
in the Fourier-conjugate representation, while their moment-order separation
decreases systematically with size.  Controlled sign masks identify the origin:
the response requires short-range sign anticorrelation aligned with the dominant
bonds, and displacing the modulation by a single site, which destroys the exactness of
that anticorrelation while leaving the sequence otherwise intact, removes the
response entirely.  An effective-dimer description
relates the alignment to the division of the onsite modulation between the
internal detuning of a bonded pair and its common-mode energy.
\end{abstract}

\begin{document}
\flushbottom
\maketitle
\thispagestyle{empty}

\section*{Introduction}

Disorder and aperiodicity provide complementary routes for controlling wave
propagation in low-dimensional systems.  In one dimension, uncorrelated
randomness generically produces Anderson localization~\cite{1}, whereas
quasiperiodic potentials can support sharply tunable localization transitions
and energy-dependent localization behavior~\cite{2,3,4,5,9,10,12,15,18,21,22,26,27,28,31,34,35,36,41,74};
comparable behavior occurs in two-dimensional quasiperiodic
models~\cite{16,17}.
These phenomena have been observed in optical media, photonic structures,
elastic waves, ultracold atoms, and superconducting-qubit platforms
~\cite{61,62,63,66,64,65,67,68,69,70}.  A particularly striking response is
reentrance: a localization diagnostic first moves toward stronger
localization, reverses over an intermediate parameter range, and finally
returns toward localization as the modulation is increased.  Such behavior
has been reported in dimerized chains, in generalized
Aubry--Andr\'e--Harper models, and in related quasiperiodic and random-matrix
settings~\cite{6,7,8,24,25,29,30,37,38,39,40,42}.
 
The interpretation of this nonmonotonic response requires care.  In a finite
chain, a state with a large but finite localization length can occupy a
substantial part of the sample and consequently acquire a large participation
ratio or generalized dimension.  These observables can therefore resemble
those of an intermediate delocalized regime even when the infinite-chain
Lyapunov exponent remains positive.  A further ambiguity is that any such
diagnostic is tied to a particular basis, so a state that appears intermediate
in real space may still be localized in its Fourier conjugate.  Establishing
the nature of a reentrant window therefore requires an asymptotic localization
diagnostic, a direct test of participation-moment saturation, and an
examination of the same states in both representations.
 
Here we study this question in a dimerized quasiperiodic chain with two sites
per unit cell.  The onsite cosine is multiplied by the deterministic
Thue--Morse (TM) sign sequence.  The TM sequence is balanced on its natural
dyadic blocks and possesses substitution correlations that are distinct from
both periodic staggering and uncorrelated randomness~\cite{45,46,47}.  It has
also appeared in a variety of condensed-matter models~\cite{57,58,59,60}.
Combined with the Su--Schrieffer--Heeger (SSH) hopping geometry~\cite{44}, it
provides a setting in which the alignment between onsite sign correlations and
the hopping hierarchy can be varied explicitly.
 
Our central result is a reentrant enhancement of the localization length.  For
intracell-dominated hopping, the window-averaged Lyapunov exponent first
increases, then falls by more than an order of magnitude, and subsequently
increases again, while remaining positive throughout; the intermediate region
is therefore a minimum of the inverse localization length rather than a zero of
the exponent.  Direct scaling across the accessible system sizes shows that
$P_2$ approaches a nonzero value at every nonzero modulation strength studied.
The finite-size variation of $D_q$ is retained as a description of how these
localized states appear on the accessible length scales, while the systematic
decrease of the moment-order separation with size rules out interpreting its
finite-size peak as a distinct phase.  The same decrease is found when the
spectral window is defined by energy instead of by eigenstate index.
Examining the same eigenstates in the Fourier-conjugate basis completes the
picture: the intermediate states are not localized in momentum space either,
but retain a structured distribution over momentum components whose scaling
depends on the moment order, whereas strongly modulated states spread broadly
over momentum as expected once they are tightly confined in real space.  These
are critical-like finite-size features: the Lyapunov exponent of the same
states is not invariant under the transformation to the dual representation,
as it is at a genuine critical point, so no critical phase is implied.  
 
The last step addresses why the response depends so sharply on the hopping
hierarchy.  The TM mask carries two structural features at once: exact sign
anticorrelation within every aligned pair of sites (a property that holds
exactly, and only when the mask and the unit cell share a common parity) and
a hierarchy of
complemented blocks on its natural dyadic scales.  We separate them using a
sequence of controlled binary masks that retain the same quasiperiodic cosine
and the same SSH geometry and change only the sign field.  Unbiased random
signs and globally balanced random signs both fail to reproduce the response,
so neither a binary sign structure nor global sign balance is by itself
sufficient.
Imposing opposite signs within each intracell pair, while randomizing all
longer-ranged order, restores the response for $J_2<J_1$ but not for
$J_2>J_1$; placing the same constraint across the intercell bonds reverses the
pattern.  What matters is therefore the alignment between the sign
anticorrelation and the bond that dominates the hopping, not the sublattice
labels themselves.  An effective-dimer estimate accounts for this.  Resolving the two onsite
energies of a bonded pair into a common-mode part and a detuning, opposite mask
signs put the smaller amplitude into the detuning and the larger into the
common-mode part, and equal signs do the reverse.  Only the detuning opposes
hybridization across the bond, the common-mode part shifting the pair rigidly,
so anticorrelation across a strong bond does not lower the modulation but moves
it to the component the bond is insensitive to.  Permuting the mask in aligned blocks
completes the picture: already the shortest aligned blocks suffice to recover
the effect, while the longer dyadic correlations of the TM sequence mainly
shift its position and strength.

\begin{figure}[h]
\includegraphics[width=80mm]{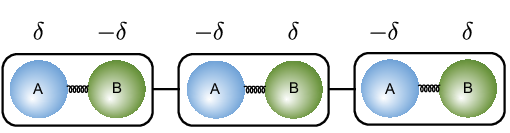}
\caption{Schematic of the TM-modulated SSH chain.  Each unit cell contains
two sublattice sites, $A$ and $B$, coupled by the intracell hopping $J_1$;
neighboring cells are coupled by $J_2$.  The displayed six-site sign mask is
the beginning of the Thue--Morse sequence.}
\label{fig:model}
\end{figure}

\section*{Model and numerical diagnostics}

We consider an open chain of $N$ unit cells and total length $L=2N$.  Its
single-particle Hamiltonian is
\begin{align}
H={}&-J_1\sum_{n=1}^{N}
\left(c_{n,A}^{\dagger}c_{n,B}+\mathrm{h.c.}\right)
-J_2\sum_{n=1}^{N-1}
\left(c_{n,B}^{\dagger}c_{n+1,A}+\mathrm{h.c.}\right) \nonumber\\
&+\sum_{n=1}^{N}\left(
\epsilon_{n,A}c_{n,A}^{\dagger}c_{n,A}
+\epsilon_{n,B}c_{n,B}^{\dagger}c_{n,B}\right),
\label{eq:H}
\end{align}
where $c_{n,A}^{\dagger}$ and $c_{n,B}^{\dagger}$ create a particle on the two
sublattices.  The onsite energies are
\begin{align}
\epsilon_{n,A}&=(-1)^{S_{2n-2}}\delta
\cos\!\left[2\pi\beta(2n-1)+\phi_1\right],\nonumber\\
\epsilon_{n,B}&=(-1)^{S_{2n-1}}\delta
\cos\!\left[2\pi\beta(2n)+\phi_2\right],
\label{eq:onsite}
\end{align}
with the TM sequence defined by
\begin{equation}
S_0=0,\qquad S_{2n}=S_n,\qquad S_{2n+1}=1-S_n.
\label{eq:TM}
\end{equation}
Thus $S_n=0110100110010110\ldots$.  Equation~\eqref{eq:onsite} aligns the mask
with the unit cell so that the two sites of a cell carry the sign indices
$2n-2$ and $2n-1$; by Eq.~\eqref{eq:TM} their signs are then exactly opposite
in every cell.  This alignment matters, and its consequences are examined
below.  We set $J_1=1$ as the energy unit,
$\beta=(\sqrt{5}-1)/2$, and $\phi_1=\phi_2=0$.  Unless otherwise stated, the
averages are taken over the central spectral sector
$m/L\in[0.475,0.525]$, where $m$ is the ordered eigenstate index.  At every
system size the complete sector is diagonalized and every eigenstate in it
enters the average; it is partitioned into forty contiguous, non-overlapping
strata, which serve as the jackknife units.  The Supplementary Information
gives the resulting geometry.

The asymptotic localization diagnostic is the Lyapunov
exponent~\cite{78,79,80,81}.  Let $t_n$
denote the hopping amplitude linking
sites $n$ and $n+1$, so that $t_n=-J_1$ on intracell bonds and $t_n=-J_2$ on
intercell bonds.  The site representation of Eq.~\eqref{eq:H} is then
\begin{equation}
t_n\psi_{n+1}+t_{n-1}\psi_{n-1}+\epsilon_n\psi_n=E\psi_n,
\label{eq:site}
\end{equation}
the amplitudes are propagated by
\begin{equation}
\begin{pmatrix}\psi_{n+1}\\ \psi_n\end{pmatrix}
=T_n(E)
\begin{pmatrix}\psi_n\\ \psi_{n-1}\end{pmatrix},\qquad
T_n(E)=
\begin{pmatrix}
(E-\epsilon_n)/t_n&-t_{n-1}/t_n\\
1&0
\end{pmatrix}.
\label{eq:transfer}
\end{equation}
The Lyapunov exponent per lattice site and localization length are
\begin{equation}
\gamma(E)=\lim_{L\rightarrow\infty}\frac{1}{L}
\ln\left\|T_L(E)\cdots T_1(E)\right\|,
\qquad \xi(E)=\gamma(E)^{-1}.
\label{eq:gamma}
\end{equation}
Products are stabilized by repeated normalization~\cite{82}.  The plotted value
$\gamma(\delta)$ is the arithmetic average over 24 approximately uniformly
selected eigenenergies in the central spectral sector.  The transfer-matrix
construction, the convergence with transfer length, the sensitivity to the
number of sampled eigenenergies, and the clean-chain calibration are given in
the Supplementary Information.

For comparison with exact diagonalization, we use the participation moments
\begin{equation}
P_q^{(i)}=\sum_{n=1}^{L}|\psi_n^{(i)}|^{2q}
\label{eq:Pq}
\end{equation}
and define the finite-size band dimension~\cite{71,73} by
\begin{equation}
\left\langle D_q(L,\delta)\right\rangle
=-\frac{\ln\left\langle P_q(L,\delta)\right\rangle}
{(q-1)\ln L}.
\label{eq:Dq}
\end{equation}
Here the spectral average is performed before the logarithm.  Reversing that order
defines the typical dimension; the two are computed from the same states, and
their largest difference anywhere in the scan is $0.017$ in $D_2$, against a
reentrant rise of $0.181$, with the minimum and the maximum at the same
$\delta$ in both.  Their difference $\Delta D$ is more sensitive to the choice,
since it subtracts two nearby quantities, and the Supplementary Information
reports both statistics throughout; the conclusions drawn below hold in each.  The comparison is tabulated in the Supplementary
Information.  In particular,
$P_2$ is the inverse participation ratio (IPR)~\cite{49,50}, while
$\mathrm{NPR}=(LP_2)^{-1}$.  For an extended state $P_2\propto L^{-1}$; for
an exponentially localized state it saturates once $L\gg\xi$.  Equation
~\eqref{eq:Dq} is used only as a finite-size diagnostic, without extrapolating
an intercept from an assumed correction law; schemes for extracting limiting
values from finite-size data are discussed in Refs.~\cite{53,56,83}.  Six
candidate correction laws are nevertheless compared in the Supplementary
Information, and their fitted coefficients and residuals are given in
Supplementary Data~2.  

\begin{figure}[h]
\includegraphics[width=\linewidth]{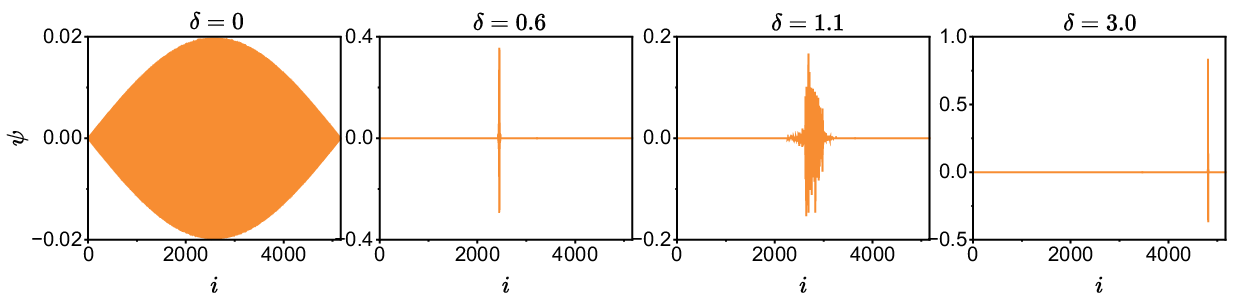}
\caption{Real-space amplitude profiles $\psi_{\ell}$ of the middle eigenstate
for $\delta=0$, $0.6$, $1.1$, and $3.0$.  The plotted quantity is the signed
amplitude, not the probability $|\psi_{\ell}|^{2}$.  The chain has $N=2584$
unit cells and $J_2=0.7$.  For even $L$ the middle state is taken as the one
with ordered index $m=L/2$, counting from $m=1$.  The profiles illustrate the large
nonmonotonic change of the occupied length scale, but do not by themselves
determine whether the intermediate state is asymptotically localized.}
\label{fig:profiles}
\end{figure}

The spectral variation quoted for $D_q$ is
obtained by a delete-one-stratum jackknife over the contiguous, non-overlapping
strata that tile the window~\cite{77}; it measures the dispersion of $D_q$
across the spectrum rather than an uncertainty on the window mean, and it does
not decrease as an independent sampling error would.  The values at every size
and modulation strength are tabulated in Supplementary Data~1.

\section*{Results}

\subsection*{Reentrant enhancement of the localization length}

Figure~\ref{fig:profiles} first illustrates the effect in real space.  The
clean mid-spectrum state spreads across the chain.  At $\delta=0.6$ it is
confined to a narrow region, while at $\delta=1.1$ it extends over a much
larger portion of the chain before becoming tightly confined again at strong
modulation.  This visual sequence motivates, but cannot settle, the
localization question because the chain is finite.

\begin{figure}[h]
\includegraphics[width=140mm]{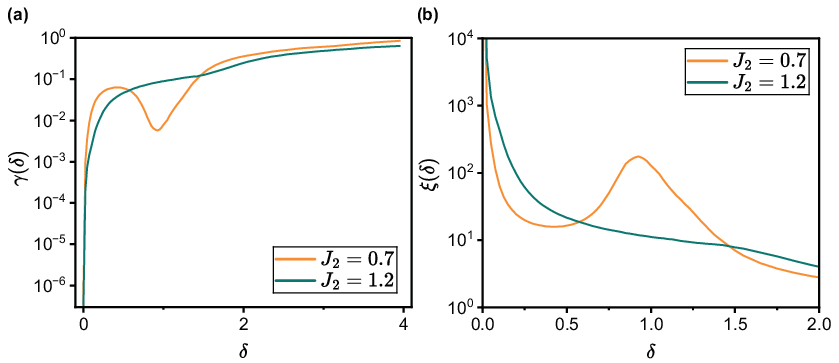}
\caption{(a) Window-averaged Lyapunov exponent $\gamma(\delta)$ and
(b) localization length $\xi(\delta)=1/\gamma(\delta)$ for $J_2=0.7$ and
$J_2=1.2$.  For $J_2=0.7$, $\gamma$ has a local maximum at
$\delta=0.425$ and a pronounced minimum at $\delta=0.925$, producing an
$11.1$-fold enhancement of $\xi$.  The $J_2=1.2$ curve is monotonic on the
same interval.}
\label{fig:gamma}
\end{figure}

The transfer-matrix result is shown in Fig.~\ref{fig:gamma}.  For $J_2=0.7$,
$\gamma$ rises from its clean value to a local maximum
$\gamma=0.06323$ at $\delta=0.425$, corresponding to $\xi=15.8$.  It then
falls to $\gamma=0.00570$ at $\delta=0.925$, where $\xi=175.5$, before
increasing again.  The intermediate increase of the localization length is
therefore a factor of $11.1$.  The small computed clean value,
$\gamma(0)=3.5\times10^{-7}$, is consistent with the exact value zero and
sets the numerical baseline.  By contrast, the response for $J_2=1.2$ is
monotonic.  Most importantly, $\gamma$ remains positive at every sampled
nonzero $\delta$ in the scan: the reentrant feature is a deep minimum of the inverse
localization length, not a zero of $\gamma$.

\begin{figure}[h]
\includegraphics[width=\linewidth]{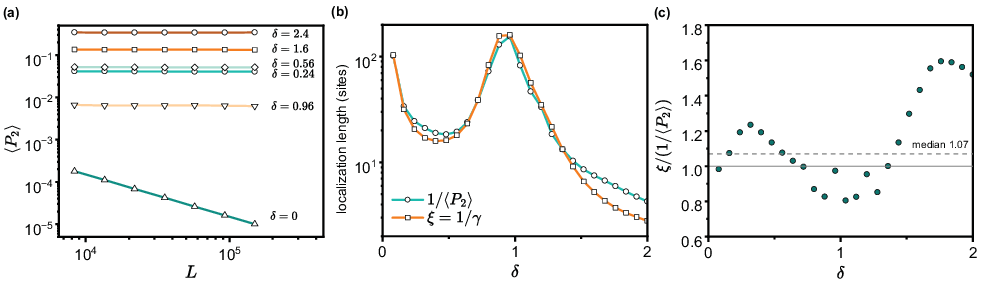}
\caption{Direct finite-size test of localization at $J_2=0.7$.
(a) The mid-spectrum average $\langle P_2\rangle$ over seven system sizes.
The clean curve follows $1/L$, whereas the curves at nonzero $\delta$
saturate.  (b) Comparison of the participation length
$1/\langle P_2\rangle$ with the transfer-matrix localization length $\xi$.
(c) Their ratio; the median value over $0<\delta<2$ is $1.07$.}
\label{fig:P2}
\end{figure}

\subsection*{Direct saturation of the participation moment}

The independent exact-diagonalization test in Fig.~\ref{fig:P2} reaches the
same conclusion without extrapolation.  The seven sizes range from
$L=8362$ to $L=150050$, a factor of $17.9$.  At $\delta=0$,
$\langle P_2\rangle$ decreases by a factor of $17.93$, as required by
$P_2\propto L^{-1}$ for an extended standing wave.  At each of the 50
nonzero values of $\delta$, however, $\langle P_2\rangle$ changes by no more
than $4.8\%$ over the same size range; the median ratio between the largest-
and smallest-size values is $1.003$.  The second participation moment has
therefore reached its localized-state plateau.

The absolute plateau is also consistent with the transfer matrix.  Because
$P_2$ depends on the entire wave-function envelope, $1/\langle P_2\rangle$
need not equal $\xi$ exactly.  Nevertheless, the two lengths follow one
another throughout the reentrant response and their median ratio is $1.07$
for $0<\delta<2$.  The nonmonotonic participation measures can consequently
be understood as the finite-size manifestation of the strongly varying, but
finite, localization length.

\subsection*{Finite-size generalized dimensions}

\begin{figure}[h]
\includegraphics[width=140mm]{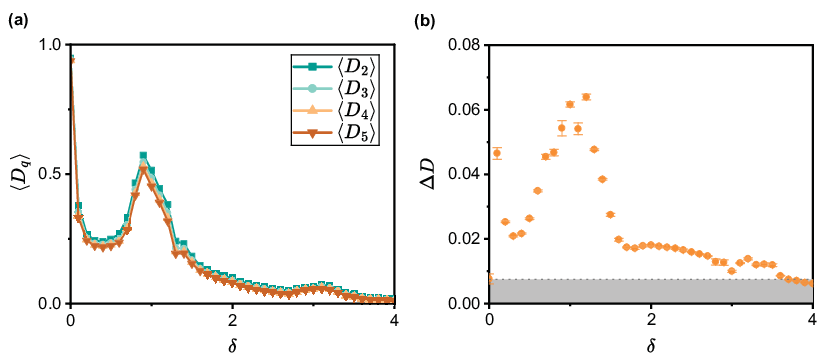}
\caption{Finite-size generalized dimensions at $J_2=0.7$ and
$L=150050$.  (a) Mid-spectrum averages $\langle D_q\rangle$ for
$q=2,3,4,5$.  (b) Moment-order separation
$\Delta D=\langle D_2\rangle-\langle D_5\rangle$.  The grey band terminates
at the exact clean finite-size value $\Delta D_{\rm clean}=0.00927$.
Vertical bars show the variation among contiguous parts of the spectral
window rather than an uncertainty on the window mean.  The complete window is
diagonalized at every size and partitioned into forty contiguous,
non-overlapping strata.}
\label{fig:Dq}
\end{figure}

Figure~\ref{fig:Dq} shows how the localized states appear on the largest
exact-diagonalization scale.  Each $D_q$ rises in the same interval in which
$\gamma$ falls and $\xi$ grows.  This coincidence is quantitative and follows
from the saturation of $P_q$.

Once $L\gg\xi$ the sum in Eq.~\eqref{eq:Pq} has effectively converged: sites
beyond the localization length carry exponentially small weight and add
nothing further, so $P_q$ becomes independent of $L$ and approaches a constant
$C_q$ fixed by the state itself~\cite{50}.  Equation~\eqref{eq:Dq} then gives
\begin{equation}
D_q(L)=-\frac{\ln C_q}{(q-1)\ln L}.
\label{eq:Dq_localized}
\end{equation}
The dimension therefore vanishes as the chain grows, but only as $1/\ln L$.
This slow approach is why a localized state can still display a large
finite-size dimension, and why widely separated sizes are needed before any
change becomes visible: the ladder used here spans a factor of $17.9$ in $L$,
yet $\ln L$ increases only from $9.03$ to $11.92$.

How $C_q$ depends on the localization length follows from the shape of the
envelope.  A state localized on a scale $\xi$ about a center $\ell_0$ can be
written as
\begin{equation}
|\psi_{\ell}|^{2}=\frac{1}{\xi}\,
f\!\left(\frac{\ell-\ell_0}{\xi}\right),
\qquad
\int f(u)\,du=1,
\label{eq:envelope}
\end{equation}
where $f$ is a normalized profile of unit width: changing $\xi$ stretches the
same profile without altering its form.  Substituting
Eq.~\eqref{eq:envelope} into Eq.~\eqref{eq:Pq} gives
\begin{equation}
C_q=c_q\,\xi^{\,1-q},
\qquad
c_q=\int f(u)^{q}\,du .
\label{eq:Cq_shape}
\end{equation}
The coefficient $c_q$ is a pure number set by the profile $f$ alone.  It
measures how peaked the envelope is and does not change when the same profile
is stretched over a larger $\xi$; for a simple exponential envelope, for
instance, $c_q=1/q$.  The whole dependence on the localization length is
therefore carried by the factor $\xi^{1-q}$.  Inserting
Eq.~\eqref{eq:Cq_shape} into Eq.~\eqref{eq:Dq_localized} gives
\begin{equation}
D_q(L)=\frac{\ln\xi}{\ln L}-\frac{\ln c_q}{(q-1)\ln L}.
\label{eq:Dq_xi}
\end{equation}
The leading term carries no $q$ dependence and is fixed by the localization
length alone, so the four dimensions collapse onto a single curve at this order
for a single state.  For the band average the collapse is exact only at $q=2$,
since $\langle P_q\rangle=\langle c_{q,i}\gamma_i^{\,q-1}\rangle$ involves the
$(q-1)$th moment of $\gamma$; the distribution of localization lengths across
the window contributes at higher moment order.  The observed hierarchy $D_2>D_3>D_4>D_5$, and the separation
\begin{equation}
\Delta D=\frac{1}{\ln L}\left[\frac{\ln c_5}{4}-\ln c_2\right],
\label{eq:dD_shape}
\end{equation}
reside in the envelope-shape term for a single state, and in that term together
with the spread of $\gamma$ across the window for the band average.  Both
contributions vanish as $1/\ln L$.

Two independent reference points fix the scale of
Eqs.~\eqref{eq:Dq_xi} and \eqref{eq:dD_shape}.  Evaluating $c_q$ for the clean
standing wave, whose envelope extends over the whole chain, reproduces the
finite-size value $\Delta D_{\rm clean}=0.11046/\ln L$ drawn as the grey band
in Fig.~\ref{fig:Dq}(b).  At the reentrant maximum the leading term gives
$\ln\xi/\ln L=0.434$ at $L=150050$ with $\xi=175.5$, against an observed peak
of $0.425$; the residual $0.009$ is the envelope-shape term of
Eq.~\eqref{eq:Dq_xi}, corresponding to $c_2=1.11$, and not an unaccounted
discrepancy.

The length entering this comparison requires a word, since the states of the
window are not identical.  We use $\xi=1/\langle\gamma\rangle$, which is
neither $\langle 1/\gamma\rangle$ nor the median localization length: at the
reentrant maximum these are $175.1$, $274.3$ and $162.0$ respectively.  For $q=2$ the arithmetic mean of $\gamma$ is nonetheless the average the
calculation requires.  Since $P_2=c_{2,i}\gamma_i$, the window average is
$\langle P_2\rangle=\langle c_{2,i}\gamma_i\rangle$, and writing this as
$c_{2,\mathrm{eff}}\langle\gamma\rangle$ with
$c_{2,\mathrm{eff}}=\langle c_{2,i}\gamma_i\rangle/\langle\gamma_i\rangle$ makes
$1/\langle\gamma\rangle$ the length that enters, exactly and without assuming the
states alike; the state dependence is carried by $c_{2,\mathrm{eff}}$ alone, and a
state-by-state check bounds the correlation it encodes at $5\%$.  For $q>2$ the same argument returns
$\langle\gamma^{q-1}\rangle$ rather than $\langle\gamma\rangle^{q-1}$, and the
two differ by $10$ to $26\%$ at $q=3$; the collapse of the leading term is
therefore exact only at $q=2$, and Eq.~\eqref{eq:Dq_xi} should be read as an
identity there and as a leading-order description at higher moment order.  The
state-resolved comparison is given in the Supplementary Information.  A large
localization volume raises $D_q$ at accessible $L$, while the dimension still
vanishes logarithmically as $L$ grows.

\begin{figure}[h]
\includegraphics[width=140mm]{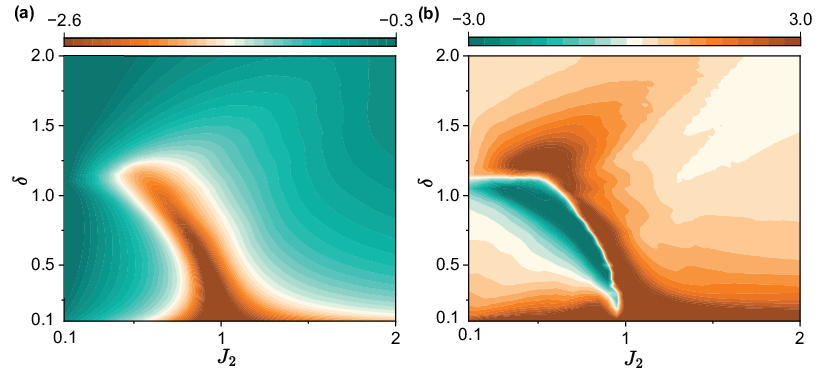}
\caption{Lyapunov response in the $(J_2,\delta)$ plane.
(a) Window-averaged Lyapunov exponent on a logarithmic scale,
$\log_{10}\gamma(J_2,\delta)$.  (b) Derivative
$\partial\log_{10}\gamma/\partial\delta$.  The low-$\gamma$ tongue in
panel (a) marks the enhancement of the localization length, while the
successive sign changes in panel (b) resolve the increase, decrease, and
subsequent increase of $\gamma$ as $\delta$ is varied.}
\label{fig:maps}
\end{figure}

The size dependence of $\Delta D$ confirms this directly: across the full
seven-size ladder it is smaller at the largest size than at the smallest at
every one of the fifty nonzero $\delta$, and decreases monotonically at
forty-nine of them.  The same holds for the typical statistic, at all fifty,
and when the central sector is defined by a fixed energy range rather than by
eigenstate index.  Changing the convention shifts $D_q$ by
about $0.005$, against a variation of about $0.17$ across the scan; the
comparison is given in the Supplementary Information.  Thus $D_q(L,\delta)$
tracks the reentrant finite-size response, whereas neither $D_q$ nor $\Delta D$
supplies evidence for an intermediate asymptotic phase.

\subsection*{Dependence on dimerization and energy}

The $J_2$ dependence in Fig.~\ref{fig:maps} connects the two representative
cuts of Fig.~\ref{fig:gamma}.  Panel (a) directly maps the asymptotic
localization diagnostic rather than a participation-based proxy.  The
low-$\gamma$ tongue occurs predominantly for $J_2<J_1$, where the intracell
bonds are dominant, and corresponds to a large but finite localization
length.  Panel (b) makes the reentrant structure explicit: a positive
$\partial\log_{10}\gamma/\partial\delta$ is followed by a negative region and
then by a return to positive slope.  For $J_2>J_1$, this sign-changing
structure is absent and $\gamma$ evolves monotonically.  The map therefore
extends the one-dimensional cuts without assigning thermodynamic phase
boundaries.

\begin{figure}[h]
\includegraphics[width=105mm]{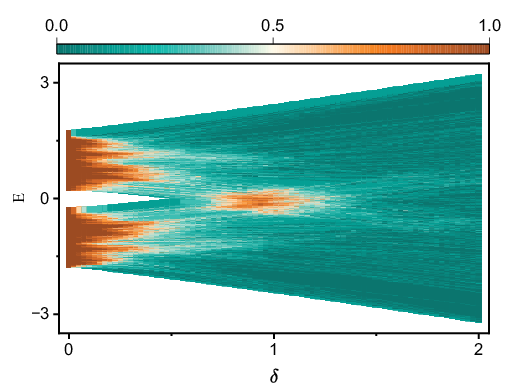}
\caption{Energy-resolved finite-size correlation dimension $D_2(E,\delta)$
at $J_2=0.7$ and $L=3\,194$ ($N=1\,597$ unit cells).  The enhancement is concentrated in the central
spectral sector and disappears again at strong modulation.}
\label{fig:energy}
\end{figure}

The energy-resolved map in Fig.~\ref{fig:energy} shows that the response is not
uniform across the spectrum.  At weak modulation, the central index sector
lies near the inner edges of the two SSH-like bands because the band center is
gapped.  As $\delta$ increases, this sector evolves toward near-zero-energy
states and exhibits the strongest increase of finite-size $D_2$.  The same
region becomes strongly localized again at larger $\delta$.  This spectral
structure motivates the central-window average used above.  Its robustness
against an energy-defined window is quantified in the Supplementary
Information, for the generalized dimensions and for the Lyapunov exponent
itself: with the window set by a fixed absolute energy interval rather than by
normalized index, the reentrant enhancement is $9.0$ against $11.1$, and at the
maximum the two conventions give localization lengths differing by $4\%$.  A
fixed-energy window cannot be made narrow and still resolve the minimum,
because the dimerization gap leaves it empty at weak modulation.

\subsection*{Dual-space characterization of the reentrant regime}

The diagnostics discussed so far characterize the structure of the eigenstates
in the site basis. For the SSH chain with two sublattices, the single-particle
eigenvalue problem corresponding to Eq.~\eqref{eq:H} can be written as
\begin{equation}
E_i \psi_{n,A}^{(i)}
=
\epsilon_{n,A}\psi_{n,A}^{(i)}
-
J_1 \psi_{n,B}^{(i)}
-
J_2 \psi_{n-1,B}^{(i)}, \qquad \text{and} \qquad E_i \psi_{n,B}^{(i)}
=
\epsilon_{n,B}\psi_{n,B}^{(i)}
-
J_1 \psi_{n,A}^{(i)}
-
J_2 \psi_{n+1,A}^{(i)} 
\end{equation}

Here, \(\psi_{n,A}^{(i)}\) and \(\psi_{n,B}^{(i)}\) are the amplitudes of the
\(i\)-th eigenstate on the two sublattices of the \(n\)-th unit cell, and
\(E_i\) is the corresponding eigenenergy. Open boundary conditions are imposed
by setting
\[
\psi_{0,B}^{(i)}=0,
\qquad
\psi_{N+1,A}^{(i)}=0 .
\]
The onsite terms \(\epsilon_{n,A}\) and \(\epsilon_{n,B}\) are determined by
the Thue--Morse-modulated quasiperiodic potential defined in
Eq.~\eqref{eq:onsite}.

Since the quantities introduced above are real-space diagnostics, they do not
by themselves exclude the possibility that an apparently intermediate state is
localized in another representation. To test this point, we analyze the same
eigenstates in momentum space. For this purpose, we map the two-sublattice
amplitudes to a single site index \(\ell=1,\ldots,L\), with \(L=2N\), according
to
\begin{equation}
\psi_{2n-1}^{(i)}=\psi_{n,A}^{(i)},
\qquad
\psi_{2n}^{(i)}=\psi_{n,B}^{(i)} .
\label{eq:sublattice_to_site}
\end{equation}
The momentum-space amplitude is then defined by the discrete Fourier transform
\begin{equation}
\tilde{\psi}_{k}^{(i)}
=
\frac{1}{\sqrt{L}}
\sum_{\ell=1}^{L}
e^{-ik\ell}
\psi_{\ell}^{(i)},
\qquad \text{where} \qquad k=\frac{2\pi m}{L},
\qquad
m=0,1,\ldots,L-1 .
\end{equation}
Although open boundary conditions are used in the diagonalization, we refer to
the Fourier-transformed amplitudes as the momentum-space representation in the
diagnostic sense. This allows us to compare the spreading of the same
eigenstate in real and momentum space.

The outcome of this comparison is stated at the outset, so that the
finite-size features described below are not read as a statement about the
thermodynamic limit.  The states of the reentrant window are exponentially
localized in real space and carry no exponential structure in the dual
representation, and their Lyapunov exponent is not invariant under the
transformation between the two representations, as it is at a genuine critical
point.  What the momentum-space diagnostics of this section display are
therefore critical-like finite-size features, not criticality.

This comparison is motivated by recent discussions of the dual-space
characterization of critical states \cite{75,76}. In this view, a critical
or multifractal state should not be exponentially localized either in real
space or in the dual momentum-space representation, and its Lyapunov exponent
is invariant under the transformation between the two.  Real-space
diagnostics alone are therefore not sufficient to establish criticality, since a state
that appears intermediate in one basis may still be localized in the other.
We analyze the same states in both representations, and test the duality
criterion quantitatively against a genuine critical state below.

We compute the generalized dimensions in both representations.  In analogy
with Eqs.~\eqref{eq:Pq} and \eqref{eq:Dq}, and at the same fixed $L$, we define
\begin{equation}
D_{q,x}^{(i)}
=
-\frac{1}{q-1}
\frac{\ln P_{q,x}^{(i)}}{\ln L},
\qquad
P_{q,x}^{(i)}
=
\sum_{\ell=1}^{L}
|\psi_{\ell}^{(i)}|^{2q},
\label{eq:Dqx}
\end{equation}
and
\begin{equation}
D_{q,k}^{(i)}
=
-\frac{1}{q-1}
\frac{\ln P_{q,k}^{(i)}}{\ln L},
\qquad
P_{q,k}^{(i)}
=
\sum_{k}
|\tilde{\psi}_{k}^{(i)}|^{2q}.
\label{eq:Dqk}
\end{equation}
As in Eq.~\eqref{eq:Dq}, the spectral average is taken before the logarithm,
and both quantities are finite-size diagnostics evaluated at a given $L$.
\begin{figure}[h]
    \includegraphics[width=\linewidth]{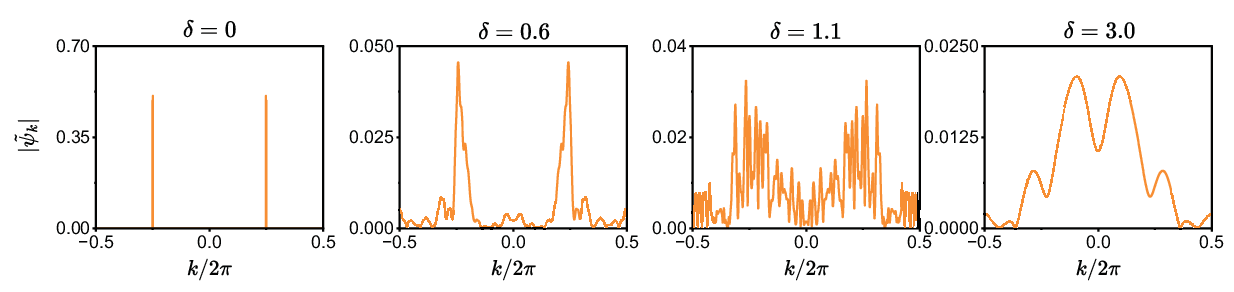}
    \caption{
    Momentum-space amplitude profiles \(|\tilde{\psi}_{k}|\) of the middle
    eigenstate for \(N=4181\), \(J_2=0.7\), and \(m/L=0.5\). The four panels show
    \(\delta=0\), \(0.6\), \(1.1\), and \(3.0\). 
}
    \label{fig:1r}
\end{figure}
For states that are localized in real space, the real-space generalized
dimensions vanish in the thermodynamic limit, \(D_{q,x}\to 0\), while the
Fourier-conjugate representation is broadly distributed and
\(D_{q,k}\to 1\). Conversely, for fully extended real-space states
\(D_{q,x}\to 1\), while the Fourier-conjugate representation is sharply peaked
and \(D_{q,k}\to 0\). A genuinely critical state is not exponentially localized
in either representation and retains nontrivial, moment-dependent dimensions in
both. Comparing \(D_{q,x}^{(i)}\) with \(D_{q,k}^{(i)}\) therefore tests
whether the states of the reentrant window are localized in one representation
or structured in both.

The comparison must be read at finite $L$, and the counting that leads to
Eq.~\eqref{eq:Dq_xi} applies equally in the Fourier-conjugate basis.  A state
confined to a length $\xi$ occupies of order $L/\xi$ momentum components, so
that
\begin{equation}
D_{q,k}(L)=1-\frac{\ln\xi}{\ln L}-\frac{\ln c_{q,k}}{(q-1)\ln L},
\label{eq:Dqk_xi}
\end{equation}
with an envelope-shape prefactor $c_{q,k}$ as before.  The leading term now
carries the localization length with the opposite sign, so a large but finite
$\xi$ raises $D_{q,x}$ and lowers $D_{q,k}$ at the same time, while the
moment-order separations $\Delta D$ and $\Delta D_k$ are both carried by the
envelope-shape terms.  Both separations vanish as $1/\ln L$, and $\Delta D_k$
decreases monotonically with $L$ at every $\delta$ examined, as $\Delta D$
does.  Structure in the momentum-space dimensions is consequently expected
wherever $\xi$ is large; it is a critical-like finite-size feature and does
not by itself indicate criticality.

The reentrant behavior is first identified from the real-space profiles in
Fig.~\ref{fig:profiles}. To obtain the corresponding momentum-space view, we
plot in Fig.~\ref{fig:1r} the amplitudes \(|\tilde{\psi}_{k}|\) of the middle
eigenstate at the same four values of \(\delta\).  The chain here is
\(N=4181\) rather than the \(N=2584\) of Fig.~\ref{fig:profiles}, so the two
figures show the corresponding state of two different chains and not the same
state twice; the comparison is between the two representations of a
mid-spectrum state at each \(\delta\), which is insensitive to that choice.

In the clean limit, \(\delta=0\), the profile consists of two sharp peaks at
opposite momenta. This is the expected structure of an open-chain standing
wave, which is formed from the superposition of \(+k\) and \(-k\) components.
When the onsite modulation is introduced, additional momentum components are
generated. At \(\delta=0.6\), the two dominant peaks remain visible but become
broadened by the potential. In the reentrant window, represented by
\(\delta=1.1\), the momentum-space profile develops a multi-component
structure, reflecting the nontrivial spatial structure of the same state in
real space. For the strongly localized state at \(\delta=3.0\), the
momentum-space amplitude becomes broad and smooth.

Thus, the momentum-space profiles provide a complementary representation of
the real-space evolution shown in Fig.~\ref{fig:profiles}. The reentrant state is not
characterized by a return to the clean two-peak momentum profile. Instead, it
shows a structured distribution over several momentum components. This
indicates that the selected real-space reentrant state is not simply localized
in momentum space. The momentum-space response also does not need to reproduce
the real-space reentrant window at exactly the same value of \(\delta\), since
real- and momentum-space participation are complementary diagnostics. This
motivates the quantitative comparison of real- and momentum-space generalized
dimensions below.
\begin{figure}[h]
    \includegraphics[width=140mm]{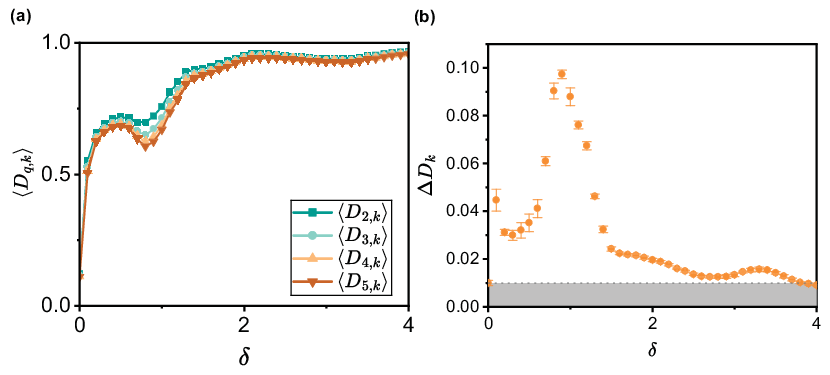}
    \caption{
    Momentum-space generalized dimensions calculated using
    eigenstates in the fixed mid-spectrum window
    $m/L\in[0.475,0.525]$ for $J_2=0.7$ and $L=8362$.
    (a) Averaged $D_{q,k}$ for $q=2,3,4,5$ as a function of the
    modulation strength $\delta$.
    (b) Momentum-space moment-order separation
    $\Delta D_k=D_{2,k}-D_{5,k}$ obtained from the same states.
    Vertical bars show the variation among contiguous parts of the spectral
    window rather than an uncertainty on the window mean.  The shaded region
    extends from zero to the clean-limit finite-size value of $\Delta D_k$.
}
    \label{fig:2r}
\end{figure}

Figure~\ref{fig:2r} gives the quantitative momentum-space
characterization of the fixed mid-spectrum window $m/L\in[0.475,0.525]$ at
$L=8362$.  Because Eqs.~\eqref{eq:Dq_xi} and \eqref{eq:Dqk_xi} both carry an explicit
$\ln L$, momentum- and real-space dimensions are to be compared at a common
size rather than against the larger-$L$ values of Fig.~\ref{fig:Dq}. The averaged $D_{q,k}$ values
generally increase with $\delta$. In the clean limit, the eigenstates
are concentrated around a small number of momentum components and the
momentum-space dimensions are relatively small. Increasing the onsite
modulation distributes the states over a larger number of momentum
components, and the dimensions approach unity in the strongly
real-space-localized regime.

The evolution is not purely monotonic.  The $D_{q,k}$ curves dip in the same
intermediate-$\delta$ region where $\xi$ is largest and $D_{q,x}$ peaks, as
Eq.~\eqref{eq:Dqk_xi} anticipates: the momentum-space minimum is the mirror
image of the real-space maximum, and both are controlled by the ratio
$\ln\xi/\ln L$.  The curves also display the hierarchy
\[
D_{2,k}>D_{3,k}>D_{4,k}>D_{5,k}.
\]
The separation between the different moment orders is summarized in
Fig.~\ref{fig:2r}(b) by
\[
\Delta D_k=D_{2,k}-D_{5,k}.
\]
The enhancement of $\Delta D_k$ in the intermediate modulation range shows
that the momentum-space dimensions acquire a pronounced $q$ dependence there.
This is the momentum-space counterpart of the critical-like finite-size
features seen in real space and, like them, it decreases systematically with
system size.
At stronger modulation, $\Delta D_k$ decreases toward its clean-limit
finite-size reference while the individual $D_{q,k}$ values approach unity,
which is the expected signature of real-space localization viewed in the
Fourier-conjugate basis.

\subsection*{Duality invariance of the Lyapunov exponent}

The dual-space criterion of Refs.~\cite{75,76} can also be applied directly to
the Lyapunov exponent rather than through the generalized dimensions.  A state
is critical when its Lyapunov exponent is invariant under the transformation
between the real-space and the dual representation, $\gamma_x=\gamma_k$; a
localized state has $\gamma_x>0$ and $\gamma_k=0$, and an extended state the
reverse.  We extract both rates from the same eigenstates with the same
estimator, namely the exponential rate of the leading descent of the amplitude
envelope away from its maximum, measured in the site index for $\gamma_x$ and
in the dual index $m$ defined by $k=2\pi\beta m$ for $\gamma_k$. The full set of values are given in the Supplementary
Information.

The estimator is calibrated on the Aubry--Andr\'e--Harper chain, which is
self-dual at $\lambda=2$ and therefore provides a genuine critical state.  At
$L=10\,946$ it returns $\gamma_x=0$ and $\gamma_k=0.724$ in the extended phase
$\lambda=1$ and $\gamma_x=0.725$ and $\gamma_k=0$ in the localized phase
$\lambda=4$, both consistent with the exact value $\ln 2 = 0.693$, and
$\gamma_x=\gamma_k=1.1\times10^{-3}$ at the critical point, where the two agree
to better than one percent and are three orders of magnitude below the
localized-phase value.  Applied to the TM-modulated chain at $J_2=0.7$ and the
same size, it returns $\gamma_k$ below $10^{-6}$ at every
modulation strength examined, including the reentrant window, while $\gamma_x$
runs from $1.2\times10^{-2}$ there to $0.60$ at $\delta=3$.  The ratio
$\gamma_k/\gamma_x$ is below $10^{-4}$ throughout, against unity for the
genuine critical state.  Duality invariance is therefore violated by four
orders of magnitude in the reentrant window, and the criterion that identifies
the critical point of the Aubry--Andr\'e--Harper chain classifies these states
as localized.

The momentum-space analysis therefore returns the same picture from the other
side.  In the reentrant region the states retain a structured distribution
over momentum components and a moment-dependent scaling, so they are not
sharply localized in the Fourier-conjugate basis; at stronger modulation they
become broadly distributed in momentum, as real-space localization requires.
Because $D_{q,x}$ and $D_{q,k}$ move in opposition, both being governed by the
same ratio $\ln\xi/\ln L$, the dual-space comparison is consistent with a
single large but finite localization length.  The states of the
reentrant window are therefore not critical in the dual-space sense set out
above: their momentum-space structure is a critical-like finite-size feature
produced by a large $\xi$ at accessible $L$, not evidence of a distinct
localization class.

\begin{figure}[h]
\includegraphics[width=120mm]{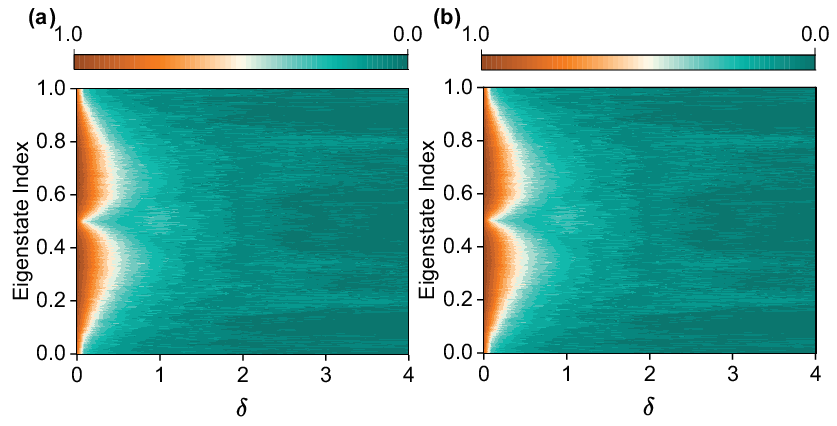}
\caption{Finite-size correlation dimension for random binary masks at
$J_2=0.7$.  (a) Ensemble-averaged $D_2$ for unbiased independent signs.
(b) Ensemble-averaged $D_2$ for a globally balanced random mask containing
equal numbers of the two signs.  In both panels, $D_2$ is shown as a function
of $\delta$ and the normalized eigenstate index $m/L$.  The system size is
$L=1974$, and each point is averaged over ten realizations.}
\label{fig:random_D2}
\end{figure}

\section*{Origin of the reentrant response}

The TM mask combines exact sign anticorrelation within every aligned pair with
a hierarchy of complemented blocks at lengths $2^p$.  We separate these two
features by comparing the TM result with random masks, pair-canceling random
ensembles, and block permutations.  All comparisons use the same
quasiperiodic cosine and SSH hopping geometry as Eq.~\eqref{eq:H}; only the
binary sign mask is changed.

Figure~\ref{fig:random_D2} shows that neither unbiased random signs nor a
globally balanced random mask reproduces the pronounced TM response.  Both
ensembles evolve essentially monotonically, and their full-spectrum
finite-size $D_2$ maps show no displaced recovery outside the selected
central window.  Binary signs and global 50/50 balance are therefore
insufficient; correlations in the sign field are required.

To isolate the shortest TM correlation, let $a_n=\pm1$ be independent random
variables and impose opposite signs within every intracell pair,
\begin{equation}
v_{2n-1}=a_n,\qquad v_{2n}=-a_n.
\label{eq:intracell_pair}
\end{equation}
This construction preserves local pair cancellation while removing the
longer-range substitution order.

\begin{figure}[h]
\includegraphics[width=140mm]{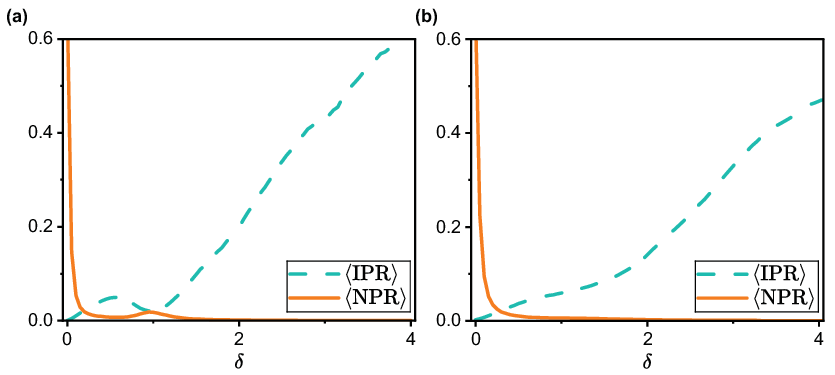}
\caption{Mid-spectrum $\langle\mathrm{IPR}\rangle$ and
$\langle\mathrm{NPR}\rangle$ for the intracell pair-canceling random
ensemble at $L=3194$.  (a) $J_2=0.7$ and (b) $J_2=1.2$.  Each point is
averaged over ten realizations of the random signs.  A nonmonotonic response appears when the sign
anticorrelation is aligned with the dominant intracell bonds.}
\label{fig:intracell}
\end{figure}

Figure~\ref{fig:intracell} displays a reentrant participation response for
$J_2=0.7<J_1$, but not for $J_2=1.2>J_1$.  The alignment can be reversed by
placing opposite signs across each intercell bond,
\begin{equation}
v_{2n}=b_n,\qquad v_{2n+1}=-b_n,
\label{eq:intercell_pair}
\end{equation}
with independent $b_n=\pm1$.  The two unpaired boundary sites under open
boundary conditions are chosen with opposite signs to maintain global
balance.

\begin{figure}[h]
\includegraphics[width=140mm]{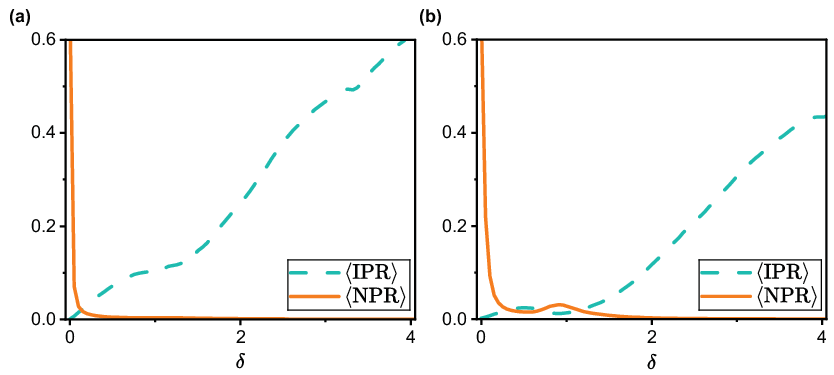}
\caption{Mid-spectrum $\langle\mathrm{IPR}\rangle$ and
$\langle\mathrm{NPR}\rangle$ for the intercell pair-canceling random
ensemble at $L=3194$.  (a) $J_2=0.7$ and (b) $J_2=1.2$.  Each point is
averaged over ten realizations of the random signs.  The response is now recovered for the
intercell-dominated hopping geometry.}
\label{fig:intercell}
\end{figure}

The crossed behavior of Figs.~\ref{fig:intracell} and
\ref{fig:intercell} demonstrates that pair cancellation is effective when it
acts across the dominant hopping bonds.  This result has a direct
effective-dimer interpretation.  Consider the two sites joined by the dominant
bond, of strength $J$, as an isolated pair with onsite energies $\epsilon_1$
and $\epsilon_2$, and resolve them into a common-mode and a differential part,
\begin{equation}
\mu_n=\frac{\epsilon_1+\epsilon_2}{2},
\qquad
\Delta_n=\frac{\epsilon_1-\epsilon_2}{2},
\label{eq:mu_delta}
\end{equation}
so that the pair has eigenvalues $\mu_n\pm\sqrt{J^2+\Delta_n^2}$.  The two
components play different roles.  The dimer-centre energy $\mu_n$ shifts both
levels rigidly and leaves the pair states unchanged; the detuning $\Delta_n$
alone opposes hybridization, splitting the bonding and antibonding
combinations further apart and rotating them away from the symmetric and
antisymmetric pair.

For two neighboring sites with cosine phases $\theta_n$ and
$\theta_n+2\pi\beta$, carrying \emph{opposite} mask signs, these are
\begin{align}
\Delta_n&=\delta a_n\cos(\theta_n+\pi\beta)\cos(\pi\beta),
\label{eq:opposite}\\
\mu_n&=\delta a_n\sin(\theta_n+\pi\beta)\sin(\pi\beta),
\label{eq:equal}
\end{align}
while for equal signs the two expressions are exchanged.  Since
$|\cos\pi\beta|\simeq0.36$ and $|\sin\pi\beta|\simeq0.93$, the amplitudes are
\begin{equation}
|\Delta_n|_{\max}=
\begin{cases}
0.36\,\delta,&v_{m+1}=-v_m,\\
0.93\,\delta,&v_{m+1}=v_m,
\end{cases}
\qquad
|\mu_n|_{\max}=
\begin{cases}
0.93\,\delta,&v_{m+1}=-v_m,\\
0.36\,\delta,&v_{m+1}=v_m,
\end{cases}
\label{eq:mismatch}
\end{equation}
and the root-mean-square values over $\theta_n$ stand in the same ratio.  The
comparison is between these amplitudes and is not a pointwise statement, since
for particular $\theta_n$ the equal-sign detuning can be the smaller of the
two.

\begin{figure}[h]
\includegraphics[width=\linewidth]{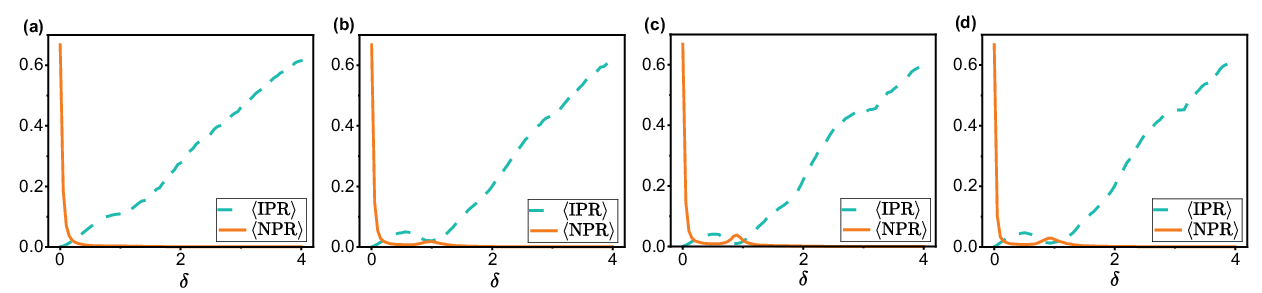}
\caption{Block-permutation test for the TM mask.  The sign field is divided
into aligned blocks of length $B=2^p$ and the blocks are randomly permuted.
The panels show the mid-spectrum participation diagnostics for (a) $p=0$,
(b) $p=1$, (c) $p=2$, and (d) $p=3$ at $L=3194$.  Each point is averaged over
ten realizations of the permutation.}
\label{fig:blocks}
\end{figure}

Anticorrelation therefore does not reduce the onsite modulation; it moves it
from the detuning into the dimer centre.  Opposite signs place the small
amplitude in $\Delta_n$, which is the component that resists hybridization
across the dominant bond, and the large amplitude in $\mu_n$, which does not
act on the pair at all.  Aligning the anticorrelation with the strong bond thus
protects the internal coherence of each dimer, at the cost of a larger spread
of dimer-centre energies from one cell to the next.  That spread is not
inconsequential: it becomes the onsite disorder of the effective chain formed
by coupling neighboring dimers through $J_2$, so the two effects act at
different levels of the problem and in opposite directions.  The estimate above
identifies which component the alignment controls; it does not by itself fix
the resulting $\delta$ dependence, which would require the full projected
inter-dimer model and is not attempted here.  What the alignment controls is
established independently, and without approximation, by the registration test
below.  For $J_2<J_1$, the TM
anticorrelation is aligned with the strong intracell bonds.  For $J_2>J_1$,
the dominant bonds are intercell bonds and the same TM constraint is
misaligned.  This accounts for the different responses in
Fig.~\ref{fig:gamma} and is consistent with the alternating modulation, which
places opposite signs across both bond types~\cite{6}.  The mechanism is
specific to the present correlated SSH chain and is not asserted to describe
all reentrant localization models.

The block-permutation test in Fig.~\ref{fig:blocks} separates the local pair
constraint from longer-range TM order.  Single-site permutation ($B=1$)
suppresses the reentrant feature.  Preserving blocks of length $B=2$ is
already sufficient to recover it, while $B=4$ and $8$ mainly change its
position and magnitude.  Short-range, bond-aligned sign anticorrelation is
therefore the enabling structural ingredient; the longer dyadic correlations
quantitatively shape the response.

\subsection*{Registration of the mask with the unit cell}

The tests above all alter the sign field.  A sharper test leaves the field
untouched and moves it: displacing the TM mask by $s$ sites relative to the SSH
unit cells changes neither the sequence, nor its correlations, nor its dyadic
block structure, but it does change which pairs of sites are aligned with the
strong bonds.

The effect is exact rather than statistical, and follows directly from
Eq.~\eqref{eq:TM}.  With the mask displaced by $s$, the two sites of cell $n$
carry the signs $(-1)^{S_{2n-2+s}}$ and $(-1)^{S_{2n-1+s}}$.  For even $s$ the
pair of indices is $(2m,2m+1)$ with $m=n-1+s/2$, so that $S_{2m}=S_m$ and
$S_{2m+1}=1-S_m$: the two signs are opposite in every cell, without exception.
For odd $s$ the pair is $(2m+1,2m+2)$, so the two signs are governed by $S_m$
and $S_{m+1}$ and the constraint is replaced by the ordinary adjacent-pair
statistics of the sequence.  Those are not the statistics of independent
variables: the Thue--Morse sequence has length-two factor frequencies
$1/6$, $1/3$, $1/3$ and $1/6$ for $00$, $01$, $10$ and $11$, so
$S_m=S_{m+1}$ with frequency $1/3$ rather than $1/2$.  Counted over
$2\times10^{5}$ cells, the two intracell signs are therefore opposite in every
cell for even $s$ and in one cell in three for odd $s$.

What the parity of the registration removes is thus not all of the
anticorrelation but its exactness: for odd $s$ a third of the cells still carry
opposite signs, and what is lost is the constraint that every cell does.
Equation~\eqref{eq:onsite} uses $s=0$.

Each construction is treated as a separate model throughout: its Hamiltonian
is diagonalized on its own, the twenty-four energies are drawn from its own
central spectral sector by the same normalized-index rule, and the transfer
exponent is evaluated at those energies.  Only the selection rule is shared, so
every $\gamma$ reported below is evaluated inside the spectrum of the operator
it describes.

\begin{figure}[h]
\includegraphics[width=80mm]{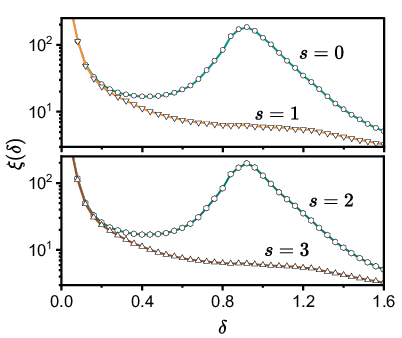}
\caption{Localization length under displacement of the Thue--Morse mask
relative to the SSH unit cells.  (a) The registration of
Eq.~\eqref{eq:onsite} ($s=0$) against a mask displaced by one site ($s=1$).
(b) Displacement by two sites ($s=2$) against three ($s=3$).  $\gamma$ is computed from the bulk
transfer matrix over $4\times10^{5}$ unit cells at $J_2=0.7$.  Each
construction is diagonalized separately and its own twenty-four eigenenergies
are selected by the normalized-index rule of Fig.~\ref{fig:gamma}.}
\label{fig:8r}
\end{figure}

Figure~\ref{fig:8r} shows the two comparisons directly, and
Table~\ref{tab:registration} collects the extrema together with the effect of
changing the cosine phase.  For even $s$ the reentrant enhancement is present
with the same extrema at $\delta=0.425$ and $\delta=0.925$ and an enhancement
factor of $11.1$ and $11.5$.  For odd $s$ it is absent: $\xi$ falls from $22.1$
at $\delta=0.20$ to $3.3$ at $\delta=1.60$, and the largest rise anywhere in
that interval is $1\%$, against a factor of eleven for the aligned chain.  The
two panels of Fig.~\ref{fig:8r} make the role of the parity explicit rather
than merely asserting it: panel (a) removes the anticorrelation by a
displacement of one site and the feature disappears, while panel (b) restores
it by a displacement of two and the feature returns at full strength.  A
displacement of seven sites behaves as one of three.  Changing the phase of the
cosine leaves the enhancement factor between $11.1$ and $11.4$ and the
positions of both extrema unchanged.

Displacing the mask index alone changes the alignment of the sign sequence
with the SSH bonds, but it also changes its alignment with the quasiperiodic
cosine, so on its own it does not isolate the bond alignment.  The control that
does is a rigid translation of the complete onsite modulation,
$\epsilon_\ell=(-1)^{S_{\ell-1+s}}\delta\cos[2\pi\beta(\ell+s)]$, which moves
the entire potential with respect to the bond pattern and leaves the relation
between the sign sequence and the cosine untouched.  It gives the same result:
enhancement factors of $11.0$ and $10.9$ for $s=0$ and $2$, and $1.01$ for
$s=1$ and $3$.  The two tests agree to better than one per cent, so the
conclusion does not depend on the residual difference between them.

\begin{table}[ht]
\centering
\caption{Reentrant enhancement of the localization length under changes of
convention.  $\xi_{\min}$ and $\xi_{\max}$ are the local extrema of
$\xi(\delta)$ bracketing the reentrant window, and the enhancement is their
ratio.}
\label{tab:registration}
\small
\begin{tabular}{lccccc}
\toprule
construction & $\delta(\xi_{\min})$ & $\xi_{\min}$ & $\delta(\xi_{\max})$
& $\xi_{\max}$ & enhancement \\
\midrule
Eq.~\eqref{eq:onsite}, $s=0$      & 0.425 & 15.82 & 0.925 & 175.9 & $11.1$ \\
mask displaced, $s=2$             & 0.425 & 15.94 & 0.925 & 182.8 & $11.5$ \\
mask displaced, $s=1$             & --    & --    & --    & --    & $1.01$ \\
mask displaced, $s=3$             & --    & --    & --    & --    & $1.01$ \\
mask displaced, $s=7$             & --    & --    & --    & --    & $1.01$ \\
cosine phase $\pi/4$              & 0.425 & 15.91 & 0.925 & 178.1 & $11.2$ \\
cosine phase $\pi/2$              & 0.425 & 15.92 & 0.925 & 181.8 & $11.4$ \\
cosine phase $\pi$                & 0.425 & 15.85 & 0.925 & 175.0 & $11.1$ \\
\bottomrule
\end{tabular}
\end{table}

\subsection*{Continuous control of the mechanism by the sublattice phase}

The above test switches the intracell anticorrelation between exact and
generic.  The relative phase of the two sublattices in Eq.~\eqref{eq:onsite} tunes it
progressively, and provides a quantitative test of the dimer picture of
Eqs.~\eqref{eq:mu_delta}--\eqref{eq:mismatch}.  With
$\phi_{\rm rel}=\phi_2-\phi_1$ the cosine arguments of the two sites of a cell
differ by $2\pi\beta+\phi_{\rm rel}$ rather than $2\pi\beta$, so that
\begin{equation}
|\Delta_n|_{\max}=\delta\left|\cos\!\left(\pi\beta+\tfrac{\phi_{\rm rel}}{2}
\right)\right|,
\qquad
|\mu_n|_{\max}=\delta\left|\sin\!\left(\pi\beta+\tfrac{\phi_{\rm rel}}{2}
\right)\right| .
\label{eq:phirel}
\end{equation}
The relative phase therefore transfers amplitude between the detuning and the
dimer-centre energy without changing anything else, and the dimer argument
predicts that the reentrant response should weaken as $|\Delta_n|$ grows toward
$|\mu_n|$ and disappear once it exceeds it.

Table~\ref{tab:phirel} shows that this is what happens.  The enhancement is
$11.0$ at $\phi_{\rm rel}=0$, where the detuning carries less than half the
amplitude of the dimer centre; it falls to $2.2$ at $\phi_{\rm rel}=\pi/4$,
where the two are nearly equal; and it is absent for $\phi_{\rm rel}=\pi/2$ and
$\pi$, where the detuning dominates.  The response is thus insensitive to a
common phase of the potential, which does not alter the decomposition, and
strongly sensitive to the relative phase, which does.  The crossover occurs
where the two amplitudes cross, which is the prediction of
Eq.~\eqref{eq:mismatch}.

\begin{table}[ht]
\centering
\caption{Reentrant enhancement as the relative sublattice phase transfers
amplitude between the internal detuning $\Delta_n$ and the dimer-centre energy
$\mu_n$, from Eq.~\eqref{eq:phirel}.  Transfer matrix over $4\times10^{5}$ unit
cells at $J_2=0.7$, with each construction diagonalized separately on a grid of
$\delta$ of spacing $0.025$.}
\label{tab:phirel}
\small
\begin{tabular}{ccccc}
\toprule
$\phi_{\rm rel}$ & $|\Delta_n|/\delta$ & $|\mu_n|/\delta$
& $|\Delta_n|/|\mu_n|$ & enhancement \\
\midrule
$0$      & 0.362 & 0.932 & 0.39 & $11.0$ \\
$\pi/4$  & 0.691 & 0.722 & 0.96 & $2.2$  \\
$\pi/2$  & 0.915 & 0.403 & 2.27 & none   \\
$\pi$    & 0.932 & 0.362 & 2.57 & none   \\
\bottomrule
\end{tabular}
\end{table}

Removing the exactness of the intracell anticorrelation, by the one minimal
change that removes it while leaving the sequence otherwise intact, therefore
eliminates the reentrant response entirely, and restoring it two sites later restores the response at
full strength.  It is worth noting how little is required: a third of the cells
still carry opposite signs for odd $s$, so it is the constraint that every
cell does, and not the presence of anticorrelated pairs as such, that the
response depends on.

\section*{Conclusion}

We have shown that the TM-modulated SSH chain exhibits a pronounced reentrant
enhancement of the localization length for intracell-dominated hopping.  The
Lyapunov exponent remains positive for every sampled nonzero modulation
strength, while its inverse grows by a factor of $11.1$ across the
intermediate window.  Direct finite-size scaling of $P_2$ shows saturation
over a factor of $17.9$ in system size, and the two length scales track one
another with a median ratio of $1.07$.  Together, these results establish that
the states remain exponentially localized even where their finite-size
participation and generalized dimensions increase sharply. The finite-size $D_q$ curves provide a useful description of this large
localization volume, but their moment-order separation decreases throughout
the seven-size ladder and does not define an additional asymptotic regime; the
same behavior is found for an energy-defined spectral window.  The reentrant
response is concentrated in the central spectral
sector and is strongest for $J_2<J_1$.  Controlled sign masks explain this hopping
dependence: anticorrelated signs move the larger part of the onsite modulation
into the dimer-centre energy, which does not oppose hybridization, and leave
only the smaller part in the detuning, which does, and they do so only when
they lie across the dominant SSH bonds.  The
registration of the mask with the unit cell makes this explicit.  Its parity
switches the intracell anticorrelation between exact and merely typical of the
sequence, and with it the reentrant response between present at full strength
and absent altogether,
while a common phase of the potential, the boundary condition and the chain
length leave the response unchanged.  The relative phase of the two sublattices
provides the same statement continuously: over the range tested it transfers
amplitude from the dimer-centre energy into the internal detuning, and the
enhancement falls as it does so, vanishing once the detuning is the larger of
the two.  The Thue--Morse hierarchy then adjusts the
position and strength of the response.
This combination of asymptotic and finite-size diagnostics distinguishes a
large, tunable localization length from a change of localization class and
provides a microscopic route for engineering nonmonotonic localization in
correlated dimerized chains.

\section*{Data availability}
The generalized dimensions at all seven system sizes and all modulation
strengths are provided as Supplementary Data~1, and the fitted coefficients,
weighted residual scores, leave-one-size-out prediction errors and per-size
residuals for every correction form as Supplementary Data~2.  The remaining
data sets generated during the current study are available from the
corresponding author on reasonable request.

\bibliography{sample}

@PREAMBLE{
 "\providecommand{\noopsort}[1]{}" 
 # "\providecommand{\singleletter}[1]{#1}%" 
}

@article{1,
  title={{Absence of Diffusion in Certain Random Lattices}},
  author={Anderson, Philip W},
  journal={Phys. Rev. },
  volume={109},
  number={5},
  pages={1492},
  year={1958},
  publisher={APS}
}

@article{2,
  title={Analyticity breaking and {Anderson} localization in incommensurate lattices},
  author={Aubry, Serge and Andr{\'e}, Gilles},
  journal={Ann. Israel Phys. Soc},
  volume={3},
  number={133},
  pages={18},
  year={1980}
}

@article{3,
  title={Single Band Motion of Conduction Electrons in a Uniform Magnetic Field},
  author={Harper, Philip George},
  journal={Proc. Phys. Soc. Sect. A },
  volume={68},
  number={10},
  pages={874},
  year={1955},
  publisher={IOP Publishing}
}

@article{4,
  title={Energy levels and wave functions of {Bloch} electrons in rational and irrational magnetic fields},
  author={Hofstadter, Douglas R},
  journal={Phys. Rev.  B},
  volume={14},
  number={6},
  pages={2239},
  year={1976},
  publisher={APS}
}

@article{5,
  title={Localization in one-dimensional incommensurate lattices beyond the {Aubry-Andr{\'e}} model},
  author={Biddle, J and Wang, B and Priour Jr, DJ and Das Sarma, S},
  journal={Phys. Rev.  A},
  volume={80},
  number={2},
  pages={021603},
  year={2009},
  publisher={APS}
}

@article{6,
  title={{Reentrant Localization Transition in a Quasiperiodic Chain}},
  author={Roy, Shilpi and Mishra, Tapan and Tanatar, Bilal and Basu, Saurabh},
  journal={Phys. Rev. Lett.},
  volume={126},
  number={10},
  pages={106803},
  year={2021},
  publisher={APS}
}

@article{7,
  title={Complete delocalization and reentrant topological transition in a {non-Hermitian} quasiperiodic lattice},
  author={Padhan, Ashirbad and Padhi, Soumya Ranjan and Mishra, Tapan},
  journal={Phys. Rev.  B},
  volume={109},
  number={2},
  pages={L020203},
  year={2024},
  publisher={APS}
}

@article{8,
  title={Phenomenon of multiple reentrant localization in a double-stranded helix with transverse electric field},
  author={Ganguly, Sudin and Sarkar, Suparna and Mondal, Kallol and Maiti, Santanu K},
  journal={Sci. Reports},
  volume={14},
  number={1},
  pages={3059},
  year={2024},
  publisher={Nature Publishing Group UK London}
}

@article{9,
  title={{Localization in One-Dimensional Lattices in the Presence of Incommensurate Potentials}},
  author={Soukoulis, CM and Economou, EN},
  journal={Phys. Rev. Lett.},
  volume={48},
  number={15},
  pages={1043},
  year={1982},
  publisher={APS}
}

@article{10,
  title={Self-dual model for one-dimensional incommensurate crystals including next-nearest-neighbor hopping, and its relation to the {Hofstadter} model},
  author={Johansson, Magnus and Riklund, Rolf},
  journal={Phys. Rev.  B},
  volume={43},
  number={16},
  pages={13468},
  year={1991},
  publisher={APS}
}

@article{12,
  title={Mobility edge and intermediate phase in one-dimensional incommensurate lattice potentials},
  author={Li, Xiao and Das Sarma, S},
  journal={Phys. Rev.  B},
  volume={101},
  number={6},
  pages={064203},
  year={2020},
  publisher={APS}
}

@article{15,
  title={Localization transitions and mobility edges in coupled {Aubry-Andr{\'e}} chains},
  author={Rossignolo, Marco and Dell'Anna, Luca},
  journal={Phys. Rev.  B},
  volume={99},
  number={5},
  pages={054211},
  year={2019},
  publisher={APS}
}

@article{16,
  title={Mixed spectra and partially extended states in a two-dimensional quasiperiodic model},
  author={Szab{\'o}, Attila and Schneider, Ulrich},
  journal={Phys. Rev.  B},
  volume={101},
  number={1},
  pages={014205},
  year={2020},
  publisher={APS}
}

@article{17,
  title={Critical states and anomalous mobility edges in two-dimensional diagonal quasicrystals},
  author={Duncan, Callum W},
  journal={Phys. Rev.  B},
  volume={109},
  number={1},
  pages={014210},
  year={2024},
  publisher={APS}
}

@article{18,
  title={One-dimensional quasicrystals with power-law hopping},
  author={Deng, X and Ray, S and Sinha, S and Shlyapnikov, GV and Santos, L},
  journal={Phys. Rev. Lett.},
  volume={123},
  number={2},
  pages={025301},
  year={2019},
  publisher={APS}
}

@article{21,
  title={The {Aubry--Andr{\'e}} model as a hobbyhorse for understanding the localization phenomenon},
  author={Dom{\'\i}nguez-Castro, GA and Paredes, R},
  journal={Eur. J. Phys.},
  volume={40},
  number={4},
  pages={045403},
  year={2019},
  publisher={IOP Publishing}
}

@article{22,
  title={{Predicted Mobility Edges in One-Dimensional Incommensurate Optical Lattices: An Exactly Solvable Model of Anderson Localization}},
  author={Biddle, J and Das Sarma, S},
  journal={Phys. Rev. Lett.},
  volume={104},
  number={7},
  pages={070601},
  year={2010},
  publisher={APS}
}

@article{24,
  title={Emergence of multiple localization transitions in a one-dimensional quasiperiodic lattice},
  author={Padhan, Ashirbad and Giri, Mrinal Kanti and Mondal, Suman and Mishra, Tapan},
  journal={Phys. Rev.  B},
  volume={105},
  number={22},
  pages={L220201},
  year={2022},
  publisher={APS}
}

@article{25,
  title={Critical analysis of the reentrant localization transition in a one-dimensional dimerized quasiperiodic lattice},
  author={Roy, Shilpi and Chattopadhyay, Sourav and Mishra, Tapan and Basu, Saurabh},
  journal={Phys. Rev.  B},
  volume={105},
  number={21},
  pages={214203},
  year={2022},
  publisher={APS}
}

@article{26,
  title={Anderson localization in one-dimensional quasiperiodic lattice models with nearest-and next-nearest-neighbor hopping},
  author={Gong, Longyan and Feng, Yan and Ding, Yougen},
  journal={Phys. Lett. A},
  volume={381},
  number={6},
  pages={588--591},
  year={2017},
  publisher={Elsevier}
}

@article{27,
  title={Fraction of delocalized eigenstates in the long-range {Aubry-Andr{\'e}-Harper} model},
  author={Roy, Nilanjan and Sharma, Auditya},
  journal={Phys. Rev.  B},
  volume={103},
  number={7},
  pages={075124},
  year={2021},
  publisher={APS}
}

@article{28,
  title={{Critical Behavior and Fractality in Shallow One-Dimensional Quasiperiodic Potentials}},
  author={Yao, Hepeng and Khoudli, Alice and Bresque, L{\'e}a and Sanchez-Palencia, Laurent},
  journal={Phys. Rev. Lett.},
  volume={123},
  number={7},
  pages={070405},
  year={2019},
  publisher={APS}
}

@article{29,
  title={Non-{Hermiticity}-induced reentrant localization in a quasiperiodic lattice},
  author={Wu, Chaohua and Fan, Jingtao and Chen, Gang and Jia, Suotang},
  journal={New J. Phys.},
  volume={23},
  number={12},
  pages={123048},
  year={2021},
  publisher={IOP Publishing}
}

@article{30,
  title={Mobility edges and reentrant localization in one-dimensional dimerized non-{Hermitian} quasiperiodic lattice},
  author={Jiang, Xiang-Ping and Qiao, Yi and Cao, Jun-Peng},
  journal={Chin. Phys. B},
  volume={30},
  number={9},
  pages={097202},
  year={2021},
  publisher={IOP Publishing}
}

@article{31,
  title={Numerical study of the localization transition of Aubry-Andr{\'e} type models},
  author={Het{\'e}nyi, Bal{\'a}zs and Balogh, Istv{\'a}n},
  journal={Phys. Rev.  B},
  volume={112},
  number={14},
  pages={144203},
  year={2025},
  publisher={APS}
}

@article{34,
  title={Electrical analogue of one-dimensional and quasi-one-dimensional {Aubry-Andr{\'e}-Harper} lattices},
  author={Ganguly, Sudin and Maiti, Santanu K},
  journal={Sci. Reports},
  volume={13},
  number={1},
  pages={13633},
  year={2023},
  publisher={Nature Publishing Group UK London}
}

@article{35,
  title={Metal-insulator transition for the almost {Mathieu} operator},
  author={Jitomirskaya, Svetlana Ya},
  journal={Ann. Math.},
  volume={150},
  pages={1159-1175},
  year={1999},
  publisher={JSTOR}
}

@article{36,
  title={{Critical Properties of Electron Eigenstates in Incommensurate Systems}},
  author={Wilkinson, M},
  journal={Proc. R. Soc. Lond. A},
  volume={391},
  number={1801},
  pages={305--350},
  year={1984},
  publisher={The Royal Society London}
}

@article{37,
  title={Coexistence of reentrant localization and dynamical delocalization in a one-dimensional non-{Hermitian} quasiperiodic lattice},
  author={Wang, Haoyu and Zheng, Xiaohong and Xiao, Liantuan and Jia, Suotang and Chen, Jun and Zhang, Lei},
  journal={Phys. Rev.  B},
  volume={112},
  number={5},
  pages={054202},
  year={2025},
  publisher={APS}
}

@article{38,
  title={Reentrant localization induced by short-range hopping in the fractal {Rosenzweig-Porter} model},
  author={Ghosh, Roopayan and Sarkar, Madhumita and Khaymovich, Ivan M},
  journal={Phys. Rev.  B},
  volume={111},
  number={22},
  pages={L220102},
  year={2025},
  publisher={APS}
}

@article{39,
  title={Emergent topological re-entrant phase transition in a generalized quasiperiodic modulated {Su-Schrieffer-Heeger} model},
  author={Wang, Xiao-Ming and Li, Shan-Zhong and Li, Zhi},
  journal={Phys. Rev.  A},
  volume={111},
  number={2},
  pages={022214},
  year={2025},
  publisher={APS}
}

@article{40,
  title={Critical analysis of multiple reentrant localization in an antiferromagnetic helix with transverse electric field: {Hopping} dimerization-free scenario},
  author={Ganguly, Sudin and Chattopadhyay, Sourav and Mondal, Kallol and Maiti, Santanu K},
  journal={SciPost Phys.},
  volume={8},
  number={1},
  pages={012},
  year={2025}
}

@article{41,
  title={Anomalous persistent current in a 1D dimerized ring with aperiodic site potential: {Non-interacting} and interacting cases},
  author={Roy, Souvik and Maiti, Santanu K and Laroze, David},
  journal={Chin. J. Phys.},
  year={2025},
  publisher={Elsevier}
}

@article{42,
  title={Investigation of reentrant localization transition in one-dimensional quasi-periodic lattice with long-range hopping},
  author={Chang, Pei-Jie and Zeng, Qi-Bo and Pi, Jinghui and Ruan, Dong and Long, Gui-Lu},
  journal={New J. Phys.},
  volume={27},
  number={5},
  pages={053501},
  year={2025},
  publisher={IOP Publishing}
}

@article{44,
  title={Solitons in {Polyacetylene}},
  author={Su, Wu-Pei and Schrieffer, John Robert and Heeger, Alan J},
  journal={Phys. Rev. Lett.},
  volume={42},
  number={25},
  pages={1698},
  year={1979},
  publisher={APS}
}

@article{45,
  title={Uber unendliche zeichenreihen},
  author={Thue, Axel},
  journal={Selsk. Skr. Mat. Nat. Kl.},
  volume={7},
  pages={1--22},
  year={1906}
}

@article{46,
  title={M{\'e}moire sur quelques relations entre les puissances des nombres},
  author={Prouhet, Eugene},
  journal={C.R. Acad. Sci. Paris},
  volume={33},
  number={225},
  pages={1851},
  year={1851}
}

@article{47,
  title={Recurrent geodesics on a surface of negative curvature},
  author={Morse, Harold Marston},
  journal={Trans. Amer. Math. Soc.},
  volume={22},
  number={1},
  pages={84--100},
  year={1921},
  publisher={JSTOR}
}

@article{49,
  title={Inverse participation ratio in 2+ $\varepsilon$ dimensions},
  author={Wegner, Franz},
  journal={Z. Phys. B },
  volume={36},
  number={3},
  pages={209--214},
  year={1980},
  publisher={Springer}
}

@article{50,
  title={Atomic vibrations in vitreous silica},
  author={Bell, R. J. and Dean, P.},
  journal={Discuss. Faraday Soc.},
  volume={50},
  pages={55-61},
  year={1970}
}

@article{53,
  title={Scaling theory of the integer quantum {Hall} effect},
  author={Huckestein, Bodo},
  journal={Rev. Mod. Phys.},
  volume={67},
  number={2},
  pages={357},
  year={1995},
  publisher={APS}
}

@article{56,
  title={A finite-size scaling analysis of the localization properties of one-dimensional quasiperiodic systems},
  author={Hashimoto, Y and Niizeki, K and Okabe, Y},
  journal={J. Phys. A},
  volume={25},
  number={20},
  pages={5211},
  year={1992},
  publisher={IOP Publishing}
}

@article{57,
  title={Structure and electronic properties of {Thue-Morse} lattices},
  author={Cheng, Zheming and Savit, Robert and Merlin, R},
  journal={Phys. Rev.  B},
  volume={37},
  number={9},
  pages={4375},
  year={1988},
  publisher={APS}
}

@article{58,
  title={Generalized {Thue-Morse} chains and their physical properties},
  author={Kol{\'a}{\v{r}}, M and Ali, MK and Nori, Franco},
  journal={Phys. Rev.  B},
  volume={43},
  number={1},
  pages={1034},
  year={1991},
  publisher={APS}
}

@article{59,
  title={Two-dimensional photonic aperiodic crystals based on {Thue-Morse} sequence},
  author={Moretti, Luigi and Mocella, Vito},
  journal={Opt. Express},
  volume={15},
  number={23},
  pages={15314--15323},
  year={2007},
  publisher={Optical Society of America}
}

@article{60,
  title={Extended and critical wave functions in a Thue-Morse chain},
  author={Ryu, CS and Oh, GY and Lee, MH},
  journal={Phys. Rev.  B},
  volume={46},
  number={9},
  pages={5162},
  year={1992},
  publisher={APS}
}

@article{61,
  title={Localization of light in a disordered medium},
  author={Wiersma, Diederik S and Bartolini, Paolo and Lagendijk, Ad and Righini, Roberto},
  journal={Nature},
  volume={390},
  number={6661},
  pages={671--673},
  year={1997},
  publisher={Nature Publishing Group UK London}
}

@article{62,
  title={Transport and Anderson localization in disordered two-dimensional photonic lattices},
  author={Schwartz, Tal and Bartal, Guy and Fishman, Shmuel and Segev, Mordechai},
  journal={Nature},
  volume={446},
  number={7131},
  pages={52--55},
  year={2007},
  publisher={Nature Publishing Group UK London}
}

@article{63,
  title={Anderson localization and nonlinearity in one-dimensional disordered photonic lattices},
  author={Lahini, Yoav and Avidan, Assaf and Pozzi, Francesca and Sorel, Marc and Morandotti, Roberto and Christodoulides, Demetrios N. and Silberberg, Yaron},
  journal={Phys. Rev. Lett.},
  volume={100},
  number={1},
  pages={013906},
  year={2008},
  publisher={APS}
}

@article{64,
  title={Localization of ultrasound in a three-dimensional elastic network},
  author={Hu, Hefei and Strybulevych, A and Page, JH and Skipetrov, Sergey E and van Tiggelen, Bart A},
  journal={Nat. Phys.},
  volume={4},
  number={12},
  pages={945--948},
  year={2008},
  publisher={Nature Publishing Group UK London}
}

@article{65,
  title={Direct observation of {Anderson} localization of matter waves in a controlled disorder},
  author={Billy, Juliette and Josse, Vincent and Zuo, Zhanchun and Bernard, Alain and Hambrecht, Ben and Lugan, Pierre and Cl{\'e}ment, David and Sanchez-Palencia, Laurent and Bouyer, Philippe and Aspect, Alain},
  journal={Nature},
  volume={453},
  number={7197},
  pages={891--894},
  year={2008},
  publisher={Nature Publishing Group}
}

@article{66,
  title={Observation of a localization transition in quasiperiodic photonic lattices},
  author={Lahini, Yoav and Pugatch, Rami and Pozzi, Francesca and Sorel, Marc and Morandotti, Roberto and Davidson, Nir and Silberberg, Yaron},
  journal={Phys. Rev. Lett.},
  volume={103},
  number={1},
  pages={013901},
  year={2009},
  publisher={APS}
}

@article{67,
  title={Observation of many-body localization of interacting fermions in a quasirandom optical lattice},
  author={Schreiber, Michael and Hodgman, Sean S and Bordia, Pranjal and L{\"u}schen, Henrik P and Fischer, Mark H and Vosk, Ronen and Altman, Ehud and Schneider, Ulrich and Bloch, Immanuel},
  journal={Science},
  volume={349},
  number={6250},
  pages={842--845},
  year={2015},
  publisher={American Association for the Advancement of Science}
}

@article{68,
  title={Single-particle mobility edge in a one-dimensional quasiperiodic optical lattice},
  author={L{\"u}schen, Henrik P and Scherg, Sebastian and Kohlert, Thomas and Schreiber, Michael and Bordia, Pranjal and Li, Xiao and Das Sarma, S and Bloch, Immanuel},
  journal={Phys. Rev. Lett.},
  volume={120},
  number={16},
  pages={160404},
  year={2018},
  publisher={APS}
}

@article{69,
  title={Anderson localization of a non-interacting {Bose-Einstein} condensate},
  author={Roati, Giacomo and D’Errico, Chiara and Fallani, Leonardo and Fattori, Marco and Fort, Chiara and Zaccanti, Matteo and Modugno, Giovanni and Modugno, Michele and Inguscio, Massimo},
  journal={Nature},
  volume={453},
  number={7197},
  pages={895--898},
  year={2008},
  publisher={Nature Publishing Group UK London}
}

@article{70,
  title={Spectroscopic signatures of localization with interacting photons in superconducting qubits},
  author={Roushan, Pedram and Neill, Charles and Tangpanitanon, J and Bastidas, Victor M and Megrant, A and Barends, Rami and Chen, Yu and Chen, Z and Chiaro, B and Dunsworth, A and others},
  journal={Science},
  volume={358},
  number={6367},
  pages={1175--1179},
  year={2017},
  publisher={American Association for the Advancement of Science}
}

@article{71,
  title={Anderson transitions},
  author={Evers, Ferdinand and Mirlin, Alexander D},
  journal={Rev. Mod. Phys.},
  volume={80},
  number={4},
  pages={1355--1417},
  year={2008},
  publisher={APS}
}

@article{73,
  title={The infinite number of generalized dimensions of fractals and strange attractors},
  author={Hentschel, H George E and Procaccia, Itamar},
  journal={Physica D},
  volume={8},
  number={3},
  pages={435--444},
  year={1983},
  publisher={Elsevier}
}

@article{74,
  title={Localization and persistent currents in a quasiperiodic disordered helical lattice},
  author={Yildiz, Taylan and Tanatar, B},
  journal={Scientific Reports},
  volume={15},
  number={1},
  pages={37307},
  year={2025},
  publisher={Nature Publishing Group UK London}
}

@article{75,
  title={Predicted critical state based on invariance of the {Lyapunov} exponent in dual spaces},
  author={Liu, Tong and Xia, Xu},
  journal={Chin. Phys. Lett.},
  volume={41},
  number={1},
  pages={017102},
  year={2024},
  publisher={Chinese Physical Society and IOP Publishing Ltd}
}

@article{76,
  title={Lyapunov Exponents as Duality-Invariant Signatures of Critical States},
  author={Liu, Tong and Xianlong, Gao},
  journal={arXiv preprint arXiv:2605.10746},
  year={2026}
}

@article{77,
  title={The jackknife and the bootstrap for general stationary observations},
  author={K{\"u}nsch, Hans R},
  journal={Ann. Stat.},
  pages={1217--1241},
  year={1989},
  publisher={JSTOR}
}

@article{78,
  title={Noncommuting random products},
  author={Furstenberg, Harry},
  journal={Trans. Amer. Math. Soc.},
  volume={108},
  number={3},
  pages={377--428},
  year={1963},
  publisher={JSTOR}
}

@article{79,
  title={A multiplicative ergodic theorem. {Lyapunov} characteristic numbers
         for dynamical systems},
  author={Oseledec, Valery I},
  journal={Trans. Moscow Math. Soc.},
  volume={19},
  pages={197--231},
  year={1968}
}

@article{80,
  title={The scaling theory of electrons in disordered solids: additional
         numerical results},
  author={MacKinnon, Angus and Kramer, Bernhard},
  journal={Z. Phys. B},
  volume={53},
  number={1},
  pages={1--13},
  year={1983},
  publisher={Springer}
}

@article{81,
  title={Localization: theory and experiment},
  author={Kramer, Bernhard and MacKinnon, Angus},
  journal={Rep. Prog. Phys.},
  volume={56},
  number={12},
  pages={1469--1564},
  year={1993},
  publisher={IOP Publishing}
}

@book{82,
  title={Products of Random Matrices in Statistical Physics},
  author={Crisanti, Andrea and Paladin, Giovanni and Vulpiani, Angelo},
  series={Springer Series in Solid-State Sciences},
  volume={104},
  publisher={Springer},
  address={Berlin},
  year={1993}
}

@article{83,
  title={Multifractal finite-size scaling and universality at the {Anderson}
         transition},
  author={Rodriguez, Alberto and Vasquez, Louella J and Slevin, Keith and
          R{\"o}mer, Rudolf A},
  journal={Phys. Rev.  B},
  volume={84},
  number={13},
  pages={134209},
  year={2011},
  publisher={APS}
}

\section*{Funding}
This work was supported by the Scientific and Technological Council of
T\"urkiye (T\"UB{\.I}TAK) under Grant No.~125F435 and the Turkish Academy of
Sciences (TUBA) under Grant No.~AD-2026.  The numerical calculations reported
in this paper were partially performed at TUBITAK ULAKBIM, High Performance
and Grid Computing Center (TRUBA resources).

\section*{Author contributions statement}
T.Y. and B.T. conceptualized the project.  T.Y. conducted the numerical
calculations and analyzed the data.  T.Y. and B.T. co-wrote the paper.

\section*{Additional information}
\textbf{Competing interests}\quad The authors declare no competing interests.

\end{document}